\documentclass[12pt]{article}
\usepackage{amssymb,graphicx}

\begin{document}
\title{Superbranes and Generic Target Space Curving}

\author{Djordje \v{S}ija\v{c}ki\thanks{email: sijacki@phy.bg.ac.yu} \\
Institute of Physics, P.O. Box 57, 11001 Belgrade, Serbia}

\date{}

\maketitle

\begin{abstract}
Embedding of a Green-Schwarz superbrane into a generic curved target space in
a general covariant way is considered. It is demonstrated explicitely, that
the customary superbrane formulation based on finite-component spinors extends
to a superspaces of restricted curving only, with the General Coordinate
Transformations realized nonlinearly over its orthogonal type
subgroups. Infinite-component, world, spinors and a recently constructed
corresponding Dirac-like equation, enable a possibility of a manifestly
covariant generic curved target space superbrane formulation.
\end{abstract}

\section{Introduction}

Quantum gravity, a theory based on General Relativity and Quantum Theory, is
still by no means one of the most outstanding problems of the contemporary
physics. There where quite a number of various attempt to tackle this problem 
without achieving a substantial breakthrough. Superstring theory, with its
subsequent superbrane based theories, is considered to be the most promising
candidate on this road. Superstring theory avoids the ultraviolet infinities
that arise in attempt to quantize gravity, it unfolds a profound way to unify
all fundamental interactions, and it strikes with its geometrical beauty and
uniqueness.

In the conventional lagrangian formulation for superbranes, the
($p+1$)-dimensional curved (locally reparametrizable) brane world volume
${\cal R}^{p+1}$ is embedded in a flat $D$-dimensional Minkowski spacetime
${\cal M}^{1,D-1}$ (Poincar\'e invariance). On the other hand, macroscopic
gravity is described classically by Einstein's theory, corresponding to a
generic curved Riemannian ${\cal R}^4$ manifold (general covariance). Thus one
is faced with an apparent difference in the manifest symmetries of these two
theories. This difference is not only of a principal nature, but is
crucial for numerous practical questions such as nonperturbative
gravitational solutions (Schwarzshild) etc. One can certainly hope to
reconstruct the full general covariance starting from the field theory
of superbranes embedded in a flat space. However, 
difficulties encountered along this line support a more pragmatic (and
in our opinion in fact the only) approach to construct an a priori
fully generally-covariant target space superbrane theory. In other
words it is desirable to construct, from the very beginning, a generic
curved target space formulation for superbranes. Having achieved this task,
one can, among other things, be in a position to study naturally the
superbrane theory on manifolds with some of the dimensions compactified. 

The aim of this paper is to study a generic curved target space symmetry of
the superbrane action, i.e. the group theoretical constraints enforced on the
action by the spinorial representations properties. As pointed out with
Y. Ne'eman, by making use of the ``there is no target to a target'' argument,
one can not embed a superstring into a generic curved target space [1]. In
this paper we show explicitely, by studying the curved target space symmetries
of a Green-Schwarz superbrane action, that the spinorial representations
properties of the $superDiff(D,R)$ group determine possible target space
curvings. There are two clearly distinguished cases: finite-dimensional and
infinite dimensional spinorial representations of the $\overline{Diff}(D,R)$
group nonlinearly realized over its orthogonal-type, e.g. $Spin(1,D-1)$, and
$\overline{SL}(D,R)$ subgroups, respectively.

\section{Bosonic brane curved space embedding}

The bosonic branes and superbranes [2] are considered in turn bellow in order
to point out their similarities and differences as for the question of a
generic curved target space embedding, as well as to fix notation and to point
out transformation properties of relevant entities.

Let us start with a bosonic p-brane embedded in a flat $D$-dimensional
Minkowski spacetime $M^{1,D-1}$. The Poincar\'e $P(1,D-1)$ group, i.e. its
homogeneous Lorentz subgroup $SO(1,D-1)$, are the physically relevant target
space symmetries, while the ($p+1$)-dimensional brane world volume is
preserved by the General Coordinate Transformation (GCT) group $SDiff(p+1,R)$.

The flat target space $p$-brane action, that permits a straightforward
transition to the supersymmetric case, is given by the following expression:
\begin{eqnarray}
&S& = \int d^{p+1}\xi \ \bigg( -\frac{1}{2p} \sqrt{-\gamma}
 \gamma^{ij} \partial_i X^m \partial_j X^n \eta_{mn}
 +\frac{p-1}{2p}\sqrt{-\gamma}  \\
&&+ \frac{1}{(p+1)!}
 \epsilon^{i_{1}i_{2}\cdots i_{p+1}}  \partial_{i_{1}}X^{m_{1}}
 \partial_{i_{2}}X^{m_{2}} \cdots  \partial_{i_{p+1}}X^{m_{p+1}}
 A_{m_{1}m_{2}\cdots m_{p+1}}(X) \bigg) , \nonumber
\end{eqnarray}
where $i$ $=$ $0,1,\dots , p$ labels the coordinates $\xi^i =
(\tau,\sigma,\rho,\dots)$ of the brane world volume with metric 
$\gamma_{ij}(\xi)$, and $\gamma=\det(\gamma_{ij})$; $m$ $=$ $0,1,\dots , D-1$ 
labels the target space coordinates $X^{m}(\xi^{i})$ with metric $\eta_{mn}$,
and $A_{m_{1}m_{2}\dots m_{p+1}}$ is a ($p + 1$)-form characterizing a
Wess-Zumino-like term in the action. 

This action can be generalized in a straightforward manner for a {\em generic
curved target space} to read in terms of the target space world variables as
follows:
\begin{eqnarray}
&S& = \int d^{p+1}\xi \
\bigg( -\frac{1}{2p}\sqrt{-\gamma}\gamma^{ij}
\partial_iX^{\tilde{m}} \partial_jX^{\tilde{n}}
g_{\tilde{m}\tilde{n}} + \frac{p-1}{2p}\sqrt{-\gamma}     \\
&&+ \frac{1}{(p+1)!} \epsilon^{i_{1}i_{2}\cdots i_{p}+1}
\partial_{i_{1}}X^{\tilde{m}_{1}}
\partial_{i_{2}}X^{\tilde{m}_{2}}\cdots
\partial_{i_{p+1}}X^{\tilde{m}_{p+1}}
A_{\tilde{m}_{1}\tilde{m}_{2}\cdots \tilde{m}_{p+1}}(X) \bigg) , \nonumber
\end{eqnarray}
where $\tilde{m}$ $=$ $0,1,\dots ,D-1$ labels the curved target space
coordinates $X^{\tilde{m}}(\xi^{i})$, with riemannian metric 
$g_{\tilde{m}\tilde{n}}$. 

The flat target space vector $X^{m}$ transforms w.r.t. a linearly realized
$D$-dimensional vectorial representation ${\cal D}^{(v)}_{SO(1,D-1)}$ of the
Lorentz group \\ $SO(1,D-1)$. The generic curved target space vector
$X^{\tilde{m}}$ transforms w.r.t. a nonlinearly realized $D$-dimensional
vectorial representation ${\cal D}^{(v)}_{Diff(D)}$ of the GCT group,
$Diff(D,R)$. As for the relevant physical subgroups, $X^{\tilde{m}}$
transforms w.r.t. a linearly realized $D$-dimensional vectorial
representation, ${\cal D}^{(v)}_{GL(D,R)}$ of the maximal linear subgroup
$GL(D,R)$ (i.e. $SL(D,R)$), as well as w.r.t. a linearly realized
$D$-dimensional vectorial representation, ${\cal D}^{(v)}_{SO(1,D-1)}$ of the
Lorentz subgroup $SO(1,D-1)$ of the $Diff(D,R)$ group. The $Diff(D,R)$ is
realized nonlinearly over $GL(D,R)$, while both $Diff(D,R)$ and $GL(D,R)$
groups are realized nonlinearly over \\ $SO(1,D-1)$).

The off-shell tensorial structure of the target space metric
$g_{\tilde{m}\tilde{n}}$ is described by a symmetric second rank irreducible
representation of the $SL(D,R)$ group, while the on-shell states are
characterized by the relevant little group, that is a subgroup of the
$SO(1,D)$ group. The off-shell, and on-shell tensor calculus is effectively
given by the $SL(D,R)$, and $SO(1,D)$ group representations, respectively,
and the $Diff(D,R)$ symmetry is nonlinearly realized over its relevant
subgroup in question.

It is a well known, and for this considerations an important fact, that
besides the scalar representations, vector representations of the $Diff(D,R)$,
$GL(D,R)$, and $SL(D,R)$ groups have the same dimensionality, $D$, as the
vector representation of the $SO(1,D-1)$ group. Due to this fact, there are
rectangular ``$D$-bines'' matrices, that connect mutually vectors of 
the above four groups. For instance, $X^{\tilde{m}} = e^{\tilde{m}}_{m} X^{m}$
connects mutually $Diff(D,R)$ and $SO(1,D-1)$ vectors taking into account the 
nonlinear realization of the $Diff(D,R)$ group over its $SO(1,D-1)$ subgroup,
etc.

\section{Superbrane curved space embedding}

The flat target space super-$p$-brane action reads [3]:
\begin{eqnarray}
&S& = \int d^{p+1} \xi 
\bigg( -\frac{1}{2p}\sqrt{-\gamma}\gamma^{ij} \Pi^m_{i} \Pi^n_{j} \eta_{mn}
+ \frac{p-1}{2p} \sqrt{-\gamma}  \\
&&+ \frac{1}{(p+1)!} \epsilon^{i_{1}i_{2}\cdots i_{p+1}}
\partial_{i_{1}}Z^{a_{1}}\partial_{i_{2}}Z^{a_{2}}\cdots
\partial_{i_{p+1}}Z^{a_{p+1}} B_{a_{1}a_{2}\cdots a_{p+1}} \bigg)\ . \nonumber
\end{eqnarray}
Here, the target space is a supermanifold with superspace coordinates \\
$Z^{a}(\xi^{i}) = (X^{m}(\xi^{i}), \Theta^{\alpha}(\xi^{i}))$, $m=0,1,\cdots
,D-1$, $\alpha = 1,2,\cdots , 2^{\left[\frac{D}{2}\right]}$. Moreover, 
\begin{eqnarray}
\Pi^{m}_{i} = \partial_{i}X^{m} + i\bar\Theta \Gamma^{m}\partial_{i}\Theta 
\end{eqnarray}
and $\Gamma^{m}$ are the corresponding $D$-dimensional target space
gamma-matrices. Note that $\Theta^{\alpha}$ transforms w.r.t. fundamental (and
its contragradient) spinorial representation of the $Spin(1,D-1)$ $\simeq$
$\overline{SO}(1,D-1)$ group, i.e. the double covering of the $SO(1,D-1)$
group. 

As for the symmetry transformation properties of the action, it is essential
that the second term in $\Pi^{m}_{i}$ transforms w.r.t. a target space
transformations as the $\partial_{i}X^{m}$ term itself. This is guaranteed,
here, by the very construction of the $D$-dimensional gamma-matrices,
$(\Gamma^{m})^{\beta}_{\alpha}$  as well as the $Spin(1,D-1)$ group  
fundamental spinorial representations properties. In particular, for
$Spin(1,D-1)$ tensor calculus, a product of a vector by a fundamental
spinorial and its contragradient representation contains these spinorial
representations upon reduction. Note, that this is not a generic feature of
Classical Lie groups/algebras, e.g. for the $SL(n,R)$ case, and a luck of this 
feature can endanger a superbrane formulation.

\subsection{Finite-component spinors}

Let us consider now the customary curved target space super-$p$-brane
action. Its derivation is based on a method inherited from supergravity and
used extensively in curving the target space in superstrings. The action is
given by the following expression [3]
\begin{eqnarray}
&S& =  \int d^{p+1}\xi \bigg( 
- \frac{1}{2p}\sqrt{-\gamma} \gamma^{ij}(\xi) E^{\tilde{m}}_i E^{\tilde{n}}_j
g_{\tilde{m}\tilde{n}} \\
&& + \frac{p-1}{2p} \sqrt{-\gamma}  +
\frac{1}{(p+1)!} \epsilon^{i_{1}i_{2}\cdots i_{p+1}}
E^{\tilde{a}_{1}}_{i_{1}} E^{\tilde{a}_{2}}_{i_{2}}\cdots
E^{\tilde{a}_{p+1}}_{i_{p+1}}
B_{\tilde{a}_{1}\tilde{a}_{2}\cdots\tilde{a}_{p+1}} \bigg)\ . \nonumber
\end{eqnarray}

Here, the target space is a supermanifold with superspace coordinates
$Z^{\tilde{a}}=(X^{\tilde{m}}, \Theta^{\alpha})$, where $\tilde{m} =
0,1,\dots ,D-1$ and $\alpha = 1,2,\dots , 2^{\left[ \frac{D}{2}
\right]}$. The number of supersymmetries, $N$, is inessential for present
discussion, and thus suppressed. Furthermore, $E^{a}_i = (\partial_i
Z^{\tilde{a}}) E^{a}_{\tilde{a}}(Z)$, where $E^{a}_{\tilde{a}}$ is the
supervielbein and $a = (m\ \alpha)$ is the so called ``tangent space
index''. As pointed in [1], there are no flat tangents to a
curved target space, which is a tangent to the brane world volume itself -
there are no frames over frames. Were it not for the spinors, generic curving
could have been achieved by replacing $X^m (\xi^i )$ by $X^{\tilde m}(\xi^i)$,
a world vector carrying finite linear representation of $SL(D,R)$ and
nonlinear representation of $Diff(D,R)$, as done in the bosonic brane case.
The $X^{\tilde m}$ transforms w.r.t. nonlinear $Diff(D,R)$ representations,
that are induced from linear representations of a formal mathematical flat
target space Lorentz group $SO(1,D-1)$. The ``tangent space index'', $a$,
labels coordinates of this formal mathematical group.

The transformation group of the bosonic part of $E^{\tilde{m}}_i$ is now the
full target space GCT group, that is nonlinearly realized under its $SL(D,R)$
subgroup. One has 
\begin{eqnarray}
superDiff(D,R)\quad \rightarrow\quad Diff(D,R)\quad\rightarrow \quad SL(D,R), 
\end{eqnarray}
where in the first step superdiffeomorphisms are mapped to
diffeomorphisms, as given by $superDiff(D,R)/Z_2$ $\simeq$ $Diff(D,R)$, and in
the second step $Diff(D,R)$ is nonlinearly realized under $SL(D,R)$. An
'effective'' transformation group of the part of $E^{\tilde{m}}_i$ that
contains spinorial variables $\Theta^{\alpha}$ is a subgroup of
$superDiff(D,R)$ that has finite-dimensional spinorial representations, and
allows for gamma-matrices construction. One has here,
\begin{eqnarray}
superDiff(D,R) \quad  \rightarrow \quad Spin(1,D-1), 
\end{eqnarray}
with superdiffeomorphisms nonlinearly realized under its $Spin(1,D-1)$
subgroup. The symmetry of the above superbrane action is an {\it intersection
of the symmetries of the terms given by even and odd variables}, i.e. an
orthogonal type subgroup of $superDiff(D,R)$. In conclusion, the mathematics
of the even part of a curved superspace yields restrictions on a possible
curving, allowing $X^{\tilde m}(\xi^i )$ to be coordinates of homogeneous
coset spaces, e.g. flat, De Sitter, anti de Sitter etc., with an orthogonal
type structure group.

\subsection{Infinite-component spinors}

Topology of the $Diff(D,R)$ group is given by the topology of its maximal
compact subgroup $SO(D)$, and thus, for $D \geq 3$, the universal covering of
the $Diff(D,R)$ group is its double covering $\overline{Diff}(D,R)$.
Likewise, the universal covering group of the $SL(D,R)$ group is its double
covering $\overline{SL}(D,R)$. It turns out, due to a way the $SO(D)$,
i.e. $SO(1,D-1)$, subgroup is embedded into $SL(D,R)$, that
$\overline{SL}(D,R)$  and $\overline{Diff}(D,R)$ groups are given by infinite
matrices [4]. There are no finite-dimensional $\overline{Diff}(D,R)$ and/or
$\overline{SL}(D,R)$ spinorial representations! The unitary
infinite-dimensional $\overline{SL}(D,R)$ spinorial representations are
constructed for various dimensions [5]. The theory of these representations
on fields is amended with a `deunitarizing automorphism'', that provides a
correct physical interpretations of the Lorentz subgroup quantities [4].

We have constructed recently a Dirac-like equation for infinite-component
spinorial $\overline{SL}(D,R)$, $D \geq 3$ fields, together with an explicit
form of the $D$-vector operator, ${\bf \Gamma^{m}}$, that generalizes Dirac's
gamma-matrices [6]. This equation is lifted up, by making use of
appropriate pseudo-frames, infinite matrices related to the quotient 
$\overline{Diff}(D,R)/\overline{SL}(D,R)$, to a fully $\overline{Diff}(D,R)$
covariant Dirac-like wave equation for a world spinor field. Moreover, the
pseudo-frames provide for a construction of generalized gamma-matrices, ${\bf
\tilde \Gamma}^{\tilde m}$ in a generic curved space.  

Let us consider now a superbrane action that reads,
\begin{eqnarray}
S = \int d^{p+1} \xi \bigg( - \frac{1}{2p}\sqrt{-\gamma} 
\gamma^{ij} \Pi^{\tilde m_{i}} \Pi^{\tilde n_{j}} g_{\tilde m\tilde n} 
+ \frac{p-1}{2p} \sqrt{-\gamma} \bigg) ,
\end{eqnarray}
where $\Pi^{\tilde m_{i}}$ is given by the following expression,
\begin{eqnarray}
\Pi^{\tilde m}_{i} = \partial_{i}X^{\tilde m} + i\bar\Theta_{\tilde\beta}
({\tilde{\bf \Gamma}}^{\tilde m})^{\tilde\beta}_{\tilde\alpha}
\partial_{i}\Theta^{\tilde\alpha}   
\end{eqnarray}
The target space is a generic curved supermanifold with superspace coordinates
$Z^{\tilde{a}}(\xi^{i}) = (X^{\tilde{m}}(\xi^{i}),
\Theta^{\tilde\alpha}(\xi^{i}))$, where $\tilde{m} = 0,1,\dots ,D-1$ and
$\tilde\alpha = 1,2,\dots , \infty$. 

This action is, by construction, invariant under the full General Coordinate
Transformations group, and thus describes a superbrane embedded into a generic
curved target space.

\section*{Acknowledgments}

This work was supported in part by MSE, Belgrade, Project-141036.

\end{document}